\documentclass[3p,twocolumn]{elsarticle} 

\usepackage{lineno}
\modulolinenumbers[0]

\usepackage[dvipsnames,table]{xcolor} 

\usepackage{hyperref}

\usepackage{xparse}

\usepackage{amsmath} 
\usepackage{amsfonts} 

\usepackage{enumitem}

\usepackage{subcaption}

\usepackage{pdfpages} 
\usepackage{pdflscape} 
\usepackage{rotating}

\usepackage{import}

\usepackage{booktabs} 

\usepackage{placeins}
\usepackage{afterpage}

\usepackage[linesnumbered,vlined,figure]{algorithm2e}
	\renewcommand{\KwSty}[1]{\textnormal{\textcolor{gray!90!black}{\ttfamily\bfseries #1}}\unskip}
	
	\SetKwComment{Comment}{\color{green!50!black}// }{}

	\newcommand{\var}{\texttt}
	\newcommand{\FuncCall}[2]{\texttt{\bfseries #1(#2)}}
	\SetKwProg{Function}{function}{}{}

\usepackage{amssymb}

\usepackage{multirow}

\usepackage{caption}

\definecolor{mpl_green}{RGB}{0, 127, 0}
\definecolor{mpl_blue}{RGB}{0, 0, 255}
\definecolor{mpl_red}{RGB}{255, 0, 0}
\definecolor{office_blue}{RGB}{0,108,209} 

\usepackage{stfloats}

\def\CNNB{$\text{CNN}_{B}$}
\def\CNNBensemble{$\text{CNN}_{B}^{ensemble}$}

\NewDocumentCommand\D{mo}{%
  \IfNoValueTF{#2}%
    {$\mathcal{D}_{\textit{#1}}$}%
    {$\mathcal{D}_{\textit{#1}}^{#2}$}%
}

\def\singlesHPO{\textit{singles-HPO}}
\def\ensemblesHPO{\textit{ensembles-HPO}}
\def\SinglesHPO{\textit{Singles-HPO}}
\def\EnsemblesHPO{\textit{Ensembles-HPO}}

\journal{Chemometrics and Intelligent Laboratory Systems}

\begin{document}


\begin{frontmatter}

\title{Automatic Neural Network Hyperparameter Optimization for Extrapolation: Lessons Learned from Visible and Near-Infrared Spectroscopy of Mango Fruit}





\author[mymainaddress]{Matthew Dirks\corref{mycorrespondingauthor}}
\ead{mcdirks@cs.ubc.ca}
\cortext[mycorrespondingauthor]{Corresponding author}

\author[mymainaddress]{David Poole}
\ead{poole@cs.ubc.ca}

\address[mymainaddress]{University of British Columbia, 2366 Main Mall, Vancouver, British Columbia, Canada}

\begin{abstract}

Neural networks are configured by choosing an architecture and hyperparameter values; 
doing so often involves expert intuition and hand-tuning to find a configuration that extrapolates well
without overfitting.
This paper considers automatic methods for configuring a neural network that extrapolates in time for the domain of visible and near-infrared (VNIR) spectroscopy. 
In particular, we study the effect of (a) selecting samples for validating configurations and (b) using ensembles.

%
Most of the time, models are built of the past to predict the future.
To encourage the neural network model to extrapolate, 
we consider validating model configurations on samples that are shifted in time similar to the test set.
We experiment with three validation set choices: 
(1) a random sample of 1/3 of non-test data (the technique used in previous work), 
(2) using the latest 1/3 (sorted by time), and 
(3) using a semantically meaningful subset of the data. 
%
%
Hyperparameter optimization relies on the validation set to estimate test-set error, 
but neural network variance obfuscates 
the true error value.
Ensemble averaging---computing the average across many neural networks---can reduce the variance of prediction errors.

%

To test these methods, we do a comprehensive study of a held-out 2018 harvest season of mango fruit given VNIR spectra from 3 prior years.
We find that ensembling improves the state-of-the-art model's variance and accuracy.
Furthermore, hyperparameter optimization experiments---with and without ensemble averaging and with each validation set choice---%
show that when ensembling is combined with 
using the latest 1/3 of samples as the validation set,
a neural network configuration is found automatically that is on par with the state-of-the-art.

\end{abstract}



\end{frontmatter}




\section{Introduction}
%

This paper considers how to automatically configure neural network hyperparameters such that it extrapolates in time for visible and near-infrared (VNIR) spectroscopy. 
Hyperparameter optimization (HPO) is a significant undertaking.
Neural networks are configured by choosing an architecture (such as number of layers) and hyperparameter values (such as learning rate),
all of which may be optimized at once during HPO.
Even when using state-of-the-art Bayesian optimization software, 
HPO still involves many decisions and intuitions
(some of which are explained in a recent tutorial \cite{Passos2022tutorial}).
This paper is about further streamlining the process of hyperparameter optimization
in order to do so automatically, without overfitting, and 
in a manner that mimics an expertly-tuned model.

A dataset is partitioned into test and non-test samples.
Given the non-test samples, the goal is to build a predictor that works the best on the test set.
The test set is only used to evaluate final models.
If a neural network is trained on all non-test data it will overfit.
To avoid overfitting, the non-test data is partitioned into calibration and validation sets 
(in machine learning literature, these are often called training and development sets).
The calibration set is used to train the model and the
validation set is used as a proxy of the test set.

Neural network hyperparameters are chosen to minimize prediction error on the validation set (in this case, it's sometimes called a tuning set).
HPO may overfit the validation set 
and the best method to combat this is an open area of research \cite{hutter2019automatedMLBook}.
To avoid overfitting, a combination of expert intuition and hand-tuning is often used. 
One approach, recently studied \cite{Mishra2021mangoes} for chemometrics, is to find stable optima where the RMSE on the validation set (with respect to the hyperparameters) is wide (doesn't change much with slight perturbations) rather than narrow \cite{hutter2019automatedMLBook}.
We take a complementary approach: 
encourage the model to extrapolate by the choice of validation samples used in hyperparameter optimization.

Extrapolating in time is often difficult because the future is different from the past.
The dataset of mangoes by Anderson et al \cite{Anderson2020mango1,Anderson2021mango2} 
is a good example:
Using spectra from 3 years, 
the goal is to predict dry matter (DM) content 
in the next year.
Thus, we want a neural network configuration that doesn't overfit the past but extrapolates well to the future.
In previous work,
the validation set is 1/3 of non-test data, sampled randomly.
We test two alternatives to avoid overfitting in HPO and encourage the neural network model to extrapolate:
First, we use the latest 1/3 of samples (sorted by time).
Second, we use a semantically meaningful subset \cite{Westad2015Validation}; specifically, the latest harvest season (2017).

Due to the stochastic nature of training algorithms,
neural networks have different weights and different errors each time they're trained.
We report the distributions of RMSE scores
for the purpose of fairly evaluating each method.
The variance of errors is also problematic for HPO because the prediction error on the validation set 
is an estimate of how well the model will perform on the test set and in deployment;
a poor estimate leads to a sub-optimal neural network configuration.

We investigate using ensembles to reduce the variance of validation-set error during hyperparameter optimization.
Ensembles (of many kinds) have been shown to improve accuracy, reduce variance, and improve robustness to domain shift \cite{Naftaly1997ensembleAvg,polikar2012ensemble,Hutter2021NeuralEnsembleSearch}.
Specifically, we obtain an ensemble by re-initializing a neural network randomly and re-training it a number of times \cite{Naftaly1997ensembleAvg,Laks2017DeepEnsembles};
this model reduces the portion of the variance that is due to random initialization.

To test these methods, we do a comprehensive study of a held-out 2018 harvest season of mango fruit
given VNIR spectra from 3 prior years \cite{Anderson2020mango1}.
We conduct hyperparameter optimization for each choice of validation set
and compare HPO with and without ensemble averaging.
The results in this study sheds light on reproducible and automated practices for configuring and training neural networks for spectroscopy;
these results can inform practitioners what steps to take in building their own models
to make predictions for future samples. 

\section{Methodologies} 

\subsection{Data set}

Visible and near-infrared (VNIR) spectra of mango fruit from four harvest seasons 
(2015, 2016, 2017, and 2018) are publicly available \cite{Anderson2020data}.
The spectral bands range $300-1100$ nm with approximately 3.3 nm intervals \cite{Anderson2020mango1}.
Near infrared spectroscopy allows for non-invasive assessment of fruit quality.
In this case, the prediction target is the percent of dry matter (DM) content. 
DM \% is an index of total carbohydrates which indicates quality of mango fruit \cite{Anderson2020mango1}.

Mishra and Passos \cite{Mishra2021mangoes} make a number of modifications to the mango fruit dataset (available online\footnote{\url{https://github.com/dario-passos/DeepLearning_for_VIS-NIR_Spectra/raw/master/notebooks/Tutorial_on_DL_optimization/datasets/mango_dm_full_outlier_removed2.mat}}), specifically: 
(1) only a subset (684--990 nm, 3.3 nm intervals) of the available spectral bands are used, 
(2) outliers have been removed from non-test-set samples,
(3) chemometric pre-processing techniques were applied and concatenated together,
and
(4) each feature is standardized separately.
Standardization of a distribution entails subtracting each value by the mean of the distribution and then dividing it by the standard deviation of the distribution.
Each sample in the dataset consists of DM \%, as the target to predict, and the concatenation of 6 vectors (each with 103 elements) which are:
\begin{enumerate}
\item The raw spectrum 
\item The first derivative of the smoothed spectrum (smoothing uses a Savitzky–Golay filter with window size of 13).
\item The second derivative of the smoothed spectrum.
\item Standardized spectrum (Standard Normal Variate, or SNV, of the spectrum).
\item The first derivative of the smoothed SNV spectrum.
\item The second derivative of the smoothed SNV spectrum.
\end{enumerate}

\subsection{Baseline Neural Network}

\begin{figure*}[htb]
    \centering
    \includegraphics[keepaspectratio,width=\linewidth]{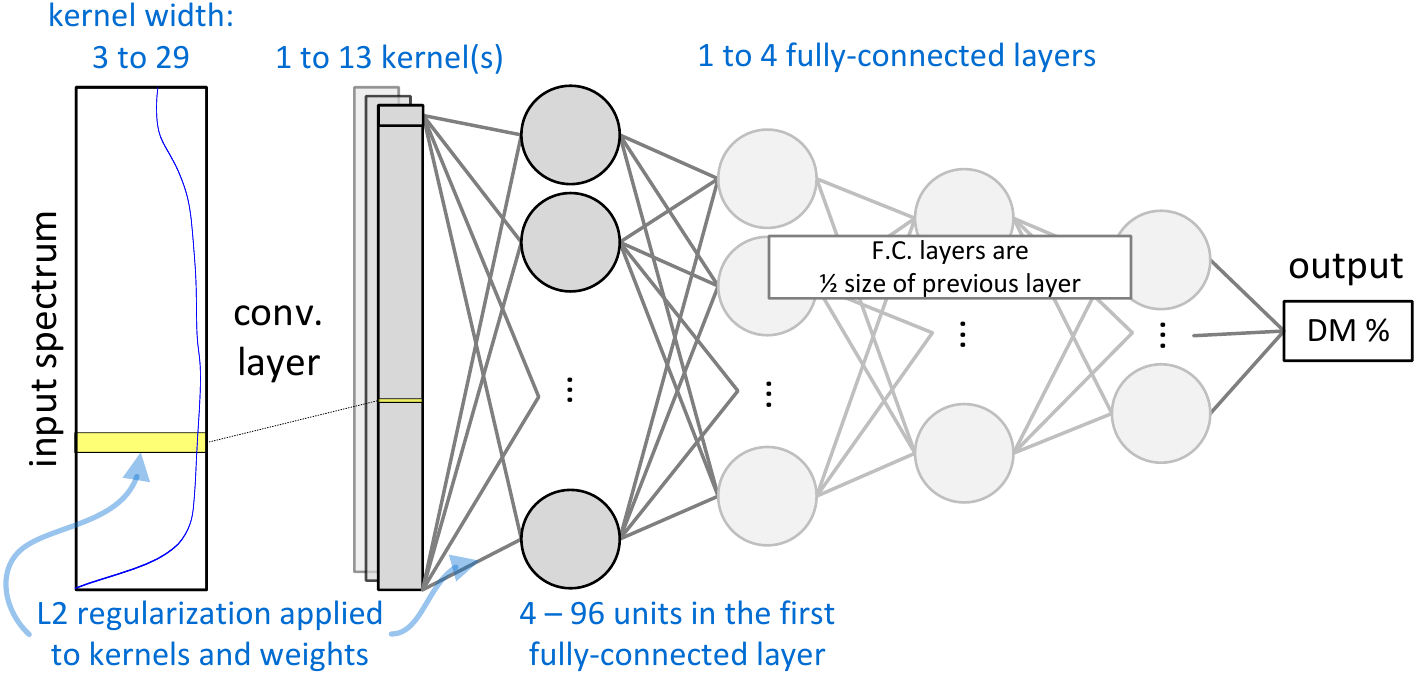}
    \caption{Neural network architecture and hyperparameter search space. Hyperparameters are shown in \textcolor{office_blue}{blue} text.}
    \label{fig:architecture}
\end{figure*}

The state-of-the-art prediction model for this dataset is a convolutional neural network (CNN),
model ``B'' by Mishra and Passos \cite{Mishra2021mangoes}, 
which we'll refer to as \CNNB{}.
This model will serve as the baseline to compare our results.
Its hyperparameters were optimized through a combination of multiple stages of grid search, expert intuition based on experience, and a careful analysis of overfitting.
For details, readers are referred to the original paper \cite{Mishra2021mangoes}
but we summarize the architecture and main hyperparameters here.
The \CNNB{} architecture (which is similar to Figure \ref{fig:architecture}) consists of a convolutional layer (1 kernel of width 21, stride 1) followed by three fully-connected layers (of size 36, 18, and 12)
with exponential linear unit (ELU) activations.
Kernel and fully-connected weights are initialized by the He normal initialization method
and regularized with an $L_2$ with a coefficient of 0.0055. 
Training proceeds for 750 epochs with mini-batches of size 128 and early stops when validation loss stops improving for 50 epochs.
Learning rate (LR) starts at 0.005 and halves each time validation loss stops improving for 25 epochs until the minimum LR is reached (1e-6).

The weights of the neural network are optimized by stochastic gradient descent using the ADAM algorithm.
Since training is stochastic (weights are initialized randomly and mini-batching randomly shuffles the data in between epochs), the weights of the network may converge to any one of many possible settings. 
In the results, we report the distribution of the errors given by randomly initializing and re-training, 
which is fairer than reporting a sample from the distribution of errors; 
this improves reproducibility and reveals more about the neural network's performance \cite{cawley2010overfitting}. 

\subsection{Hyperparameter Search Space and Optimization}
\label{sec:HPO}

Neural networks can take on many different architectures, each with many possible hyperparameters.
Any specific assignment of all the hyperparameters is referred to as a configuration.
In this study, the problem of neural architecture search is treated as additional hyperparameters and will be optimized in conjunction with other hyperparameters \cite{Hutter2018autoDL}.

The space of possible architectures used in our hyperparameter search is
based on typical architectures used in spectroscopy applications \cite{acquarelli2017cnnForVib,lui2017CNNRaman,malek20181dCNN,fearn2018modernCNN,fan2019raman,chatzidakis2019sciReport,zhang2019DeepSpectra,yang2019progressAndGuide}
and \CNNB{} \cite{Mishra2021mangoes} in particular.
These neural network architectures all follow a similar pattern: one or more convolutional layers followed by one or more fully-connected layers, with an activation function after each layer.
Our architecture and hyperparameter search space is visualized in Figure \ref{fig:architecture}).
%
%
%
The particular architecture and search space we use is somewhat arbitrary
because
the goal of the experiments is to test the HPO method under different conditions.

The search space consists of 5 hyperparameters, and we deliberately structure them to leave out conditional hyperparameters to keep it simple.
More hyperparameters may also be included, with the potential of increasing prediction accuracy, 
but would require more computation and for the purpose of testing HPO it is not necessary. 
The hyperparameters and their ranges are listed in Table \ref{tbl:hyperparams} (ranges based on published literature and standard practice).
All fully-connected layers, other than the first one, are half the size of the previous layer;
this keeps the search space smaller and simpler.
For example, when there are 3 fully-connected layers
and the first layer has 36 units (as in \CNNB{}),
then the second and third layers will have 18 and 9 units.
Search space for the convolution kernel's width is approximately logarithmic and rounded to the nearest odd number, ranging from 3 to 29.
All hyperparameter values are integers except $L_2$ regularization which is a floating-point number.

\begin{table}[htb]
\centering
\begin{tabular}{ll}
\hline
Hyperparameter                    & Range                                \\ \hline
$L_2$ coefficient                 & {[}$10^{-4}$, 1{]}                   \\
Number of kernels                 & {[}1, 13{]}                          \\
Kernel width                      & 3,5,7,11,15,21,29                    \\
Number of FC layers               & {[}1, 4{]}                           \\
Number of FC units                & {[}4, 96{]} (step 4)                 \\ \bottomrule
\end{tabular}
\caption{Hyperparameters' search space.}
\label{tbl:hyperparams}
\end{table}

In the field of automated machine learning (Auto-ML), the state-of-the-art for automatically searching through a hyperparameter search space is Bayesian optimization.
Bayesian optimization (BO) for HPO \cite{Bergstra2015hyperoptTutorial,deFreitas2016ReviewBO,Passos2021HPOwheat} chooses which hyperparameter configuration to try next.
The first trial configuration is sampled randomly.
Subsequent configurations are selected based on a mix of how well they are estimated to perform and how uncertain the estimate is.
We use an implementation of Bayesian optimization called hyperopt \cite{Bergstra2013hyperopt}.
We follow standard practice in automatic hyperparameter optimization (HPO);
readers are referred to a tutorial on HPO for spectral modelling to learn how to perform similar experiments from scratch
\cite{Passos2022tutorial}.



\newcommand{\colorDrand}[1]{{\color{BurntOrange}#1}}
\newcommand{\colorDval}[1]{{\textcolor{office_blue}{#1}}}

\begin{algorithm}

    \Function{train(config, \D{split})}{
        \var{NN} with \var{config}, initialize randomly\\                        \label{line:NN-init}
        Train \var{NN} weights on \D{split}[cal], \\                              \label{line:NN-train}
        with ES and LR schedule based on \D{split}[\colorDval{val}]\\             \label{line:NN-ES-LR}
        \Return{NN}\\
    }
    \Function{HPO(\D{split})}{ \label{line:HPO-function}
        \For{config \KwSty{in} configurations}{                                   \label{line:HPO-BO}
            \var{NN} = \FuncCall{train}{config, \D{split}}\\                            \label{line:HPO-train-NN} 
            \var{RMSE}$_{config}$ = \FuncCall{NN.predict}{\D{split}[\colorDval{val}]}\\            \label{line:HPO-score}
        }
        \Return{config} with lowest RMSE\\                                        \label{line:HPO-return}
    }

    \Function{final\_evaluation(\D{split})}{
        \var{final\_config} = \FuncCall{HPO}{\D{split}}\\                          \label{line:eval-HPO}
        \var{final\_NN} = \FuncCall{train}{best\_config, \colorDrand{\D{rand}}}\\  \label{line:eval-retrain}
        \var{RMSE} = \FuncCall{best\_NN.predict}{\D{}[test]}\\                    \label{line:eval-score}
    }

    \caption{
        Experiment overview (without ensembling). 
        Pseudo-code of neural network training (\var{train} function) and
        hyperparameter optimization (\var{HPO} function) 
        showing how validation set choices are used 
        and how RMSE is calculated in \var{HPO} and \var{final\_evaluation}.
        \D{split} is the dataset that has been partitioned (\textit{split}) in some way (described in Section \ref{sec:val}).
        \D{split}[val] is the validation set 
        and \D{split}[cal] is the calibration set for this \textit{split}.
        %
    }
    \label{code}
\end{algorithm}

Pseudo-code in Figure \ref{code}, line \ref{line:HPO-function}, shows the general idea: 
select a hyperparameter configuration ($\var{config}$) to try next via Bayesian optimization (line \ref{line:HPO-BO}),
train a neural network using that configuration (line \ref{line:HPO-train-NN}),
obtain a score (RMSE) for this configuration (line \ref{line:HPO-score}) using predictions on a validation set (validation set described in next section),
and, finally, choose the configuration with the lowest RMSE (line \ref{line:HPO-return}).
We stop HPO after completing approximately 2000 configurations 
(this is approximate as many were run in parallel and we stopped HPO manually once there was greater than 2000).

\subsection{Choice of Validation Set}
\label{sec:val}

\begin{table*}[htb]
\centering
\begin{tabular}{@{}llll@{}}
\toprule
      & Size of calibration set & Size of validation set & Validation set is... \\ \midrule
\D{rand}  & 6642            & 3272  & a random 33\% as in previous work \cite{Anderson2020mango1}. \\
\D{shift} & 6642            & 3272  & the latest 33\% (sorted by time). \\
\D{2017}  & 5045            & 4869  & the latest harvest season (2017). \\ \bottomrule
\end{tabular}
\caption{Validation set choices and number of data points in validation set and corresponding calibration set. Test set is fixed with a size of 1448.}
\label{tbl:val-sets}
\end{table*}

Pseudo-code in Figure \ref{code} shows where the calibration, validation, and test sets are used in training and HPO.
A calibration set (\textit{cal}) is used for training the neural network weights (line \ref{line:NN-train})
and a validation set (\textit{\colorDval{val}}) is used to inform the training procedure when to early-stop (ES) and 
when to drop the learning rate (LR)%
\footnote{The epoch at which early stopping occurs and the LR schedule (epochs at which to drop the LR) 
could instead be learned hyperparameters, 
but standard practice is to use the validation set to dynamically adapt as the training proceeds.
We adopted this because we are only testing HPO with a limited set of hyperparameters.}
 (line \ref{line:NN-ES-LR}).
The validation set (\textit{\colorDval{val}}) is also used in HPO to score configurations (line \ref{line:HPO-score}).
The test set, \D{}[test], is used in the final evaluation to score how well HPO performed (line \ref{line:eval-score}).
Test set is described next
followed by 3 choices for partitioning samples into calibration and validation sets.

As in previous works \cite{Anderson2020mango1,Anderson2021mango2,Mishra2021mangoes}, the 2018 harvest season is used as the test set for evaluation.
For a predictive model in the fruit industry to be practical, it must be robust to year-to-year variations. 
Since the hope is that the model can extrapolate to a new season,
a whole season is the right choice for the test set 
 \cite{Westad2015Validation}.
%
%

The samples from the 3 years prior (2015, 2016, 2017) are partitioned into calibration and validation sets.
In previous work \cite{Mishra2021mangoes} (and standard practice), the validation set is 33\% of the 
the training data, randomly sampled without replacement (which we'll call \D{rand}[val]), 
and the calibration set (\D{rand}[cal]) is the remaining 67\%. 
Samples can also be chosen more deliberately, keeping in mind the goal to extrapolate.
We test two other choices:

\begin{enumerate}
    \item \D{shift}: 
        Sort samples by date and split somewhere in the middle, so that the validation set is later than the calibration set which then requires the model to extrapolate in time in order to do well.
        We use the last 33\% of samples for validation and the remainder for calibration;
        this choice keeps the size of the sets equal to previous work.

    \item \D{2017}: 
        Select an entire season to force the model to extrapolate to a future season, just like we expect it do for the test set.
        Since the test set is from 2018, we use the 2017 harvest season samples
        (harvest seasons do not follow the calendar year but samples are labelled with their season in the dataset \cite{Anderson2020data}).
        2017 samples comprise 49\% of non-test data which is considerably more validation samples than before.
\end{enumerate}

\begin{figure}[htb]
    \centering
    \includegraphics[keepaspectratio,width=1\linewidth]{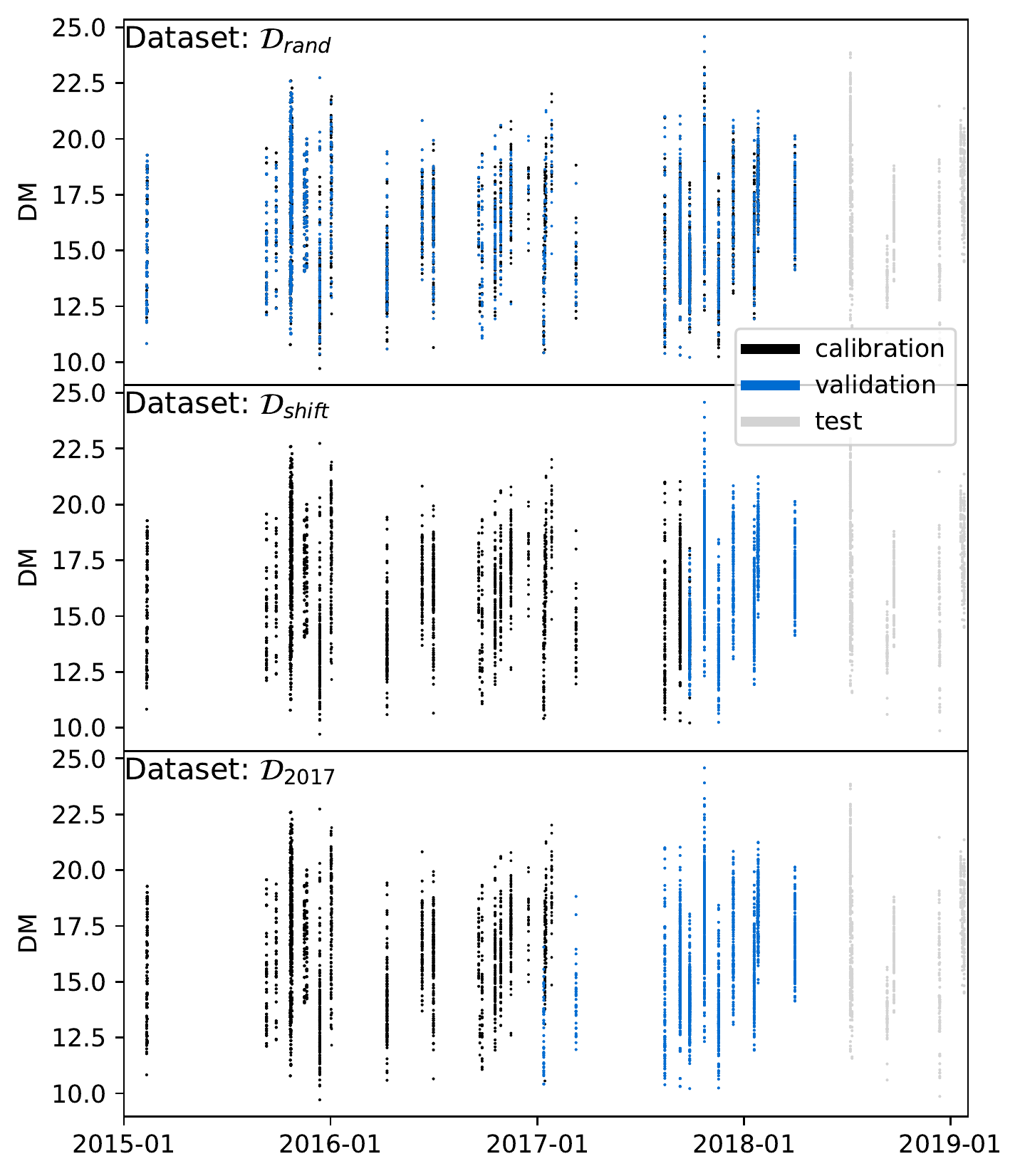}
    \caption{Calibration, validation, and test set for each partitioning of the dataset: \D{rand}, \D{shift}, and \D{2017}.
             Each point is a sample from the dataset colored by assignment to calibration, validation, or test set.
             DM \% is plotted on the y-axis with the date the sample was collected on the x-axis.
            }
    \label{fig:data-splits}
\end{figure}

Validation set choices, and the sizes of each set, are summarized in Table \ref{tbl:val-sets}.
Figure \ref{fig:data-splits} visualizes these choices by showing the date and DM \% of each sample in each set.
The $\var{final\_evaluation}$ function (in pseudo-code Figure \ref{code}) is run on each validation set choice.
That is,
\D{split} is one of \D{rand}, \D{shift}, or \D{2017}.
\colorDrand{\D{rand}} is always used to test the final models regardless of which \D{split} is used for HPO (line \ref{line:eval-retrain}).

\subsection{Neural Network Ensemble Averaging}

Neural networks are trained by minimizing an objective function,
but there can be many sub-optimal weight settings after training
depending on a number of factors including
    how the weights are initialized
    and the order of samples seen during training.
As we show in the results, a neural network's prediction score varies under different random initializations
(repeating the $\var{train}$ function in pseudo-code Figure \ref{code} results in neural networks with different weights and different prediction errors).
As datasets get bigger, the variance is reduced \cite{Dietterich2000Ensembles}, but when that's unavailable, as is often the case, 
another method is to use ensembles \cite{polikar2012ensemble}.

An ensemble is when multiple models' predictions are combined to produce a final prediction. 
Ensembles are able to efficiently utilize these models
by smoothing predictions 
which reduces overfitting
\cite{cooper1993whenNetworksDisagree}.
There are different strategies for choosing models and for combining their predictions \cite{Dietterich2000Ensembles,polikar2012ensemble,Hutter2021NeuralEnsembleSearch}.

A straightforward, yet powerful, strategy is to train a neural network multiple times with different random seeds (random initializations)
then average their predictions \cite{Naftaly1997ensembleAvg}, which is the approach used in this paper (sometimes called ensemble averaging over initial conditions).
Much more sophisticated ensemble methods exist \cite{polikar2012ensemble}, 
but recent literature suggests they don't live up to expectations \cite{ashukha2020pitfalls}.
A resurgence of recent studies, under the name Deep Ensembles \cite{Laks2017DeepEnsembles}, continues to demonstrate the effectiveness of this approach on modern deep neural networks;
optimization from random initializations explores different modes leading to diverse ensemble members and, therefore, improved accuracy and out-of-distribution robustness \cite{Fort2019EnsemblesLossLandscape}.
%
%

\subsection{Experiment}

We conduct hyperparameter optimization (HPO) for 6 scenarios:
for each validation set choice (\D{rand}, \D{shift}, and \D{2017}),
and with and without ensembling.
The final models from each HPO scenario are evaluated and compared.

\subsubsection{HPO With and Without Ensembling}
HPO without ensembling (referred to as \singlesHPO{}) follows the process laid out in the pseudo-code (Figure \ref{code}).
HPO with ensembling (referred to as \ensemblesHPO{}) creates an ensemble for each configuration.
In Figure \ref{code}, line \ref{line:HPO-train-NN} is repeated a number of times to make the ensemble's models.
Each model in the ensemble makes predictions on the validation set, as in line \ref{line:HPO-score}. 
The predictions from the models are averaged, and RMSE for the ensemble is calculated using the averaged predictions.

\subsubsection{Final Models With and Without Ensembling}
For each HPO scenario, 
the final configuration from HPO 
(Figure \ref{code}, line \ref{line:eval-HPO})
is evaluated.
We evaluate the baseline (\CNNB{}) configuration and final configurations from \singlesHPO{} scenarios 
both as single models and as ensemble models.
Since \ensemblesHPO{} scenarios already utilize ensembles, 
we evaluate their final configurations as ensemble models only.

For single models,
the final model ($\var{final\_NN}$, line \ref{line:eval-retrain})
is obtained by training on the 
original partitioning of the dataset (\colorDrand{\D{rand}}).
The final model makes predictions on the test set (\D{}[test]).
Then a distribution of RMSE scores is generated by repeatedly training and testing (repeating lines \ref{line:eval-retrain}-\ref{line:eval-score}).

For ensemble models,
an ensemble is created by re-training (repeating line \ref{line:eval-retrain}),
then the final RMSE is calculated using the average of the ensemble's models' predictions on the test set.
A distribution of RMSE scores is again generated, this time by re-training the entire ensemble a number of times.

\subsubsection{Computing Software Environment}
TensorFlow 2.6 and Python 3.9 software is used to train the neural networks.
Experiments are run on a high-performance compute cluster named Cedar, 
hosted by Simon Fraser University for Compute Canada and Digital Research Alliance of Canada,
with NVIDIA V100 and P100 GPUs and Intel Xeon E5-2650 2.2GHz and Intel Xeon Silver 4216 2.1GHz CPUs
(1352 GPU devices and 94528 CPU cores).


\section{Results and Discussion}
\label{sec:results}

Experimental results are laid out in Figure \ref{fig:HPO_boxplot_groups};
we explain each panel in-turn in the following sections.
First we report the distributions of single models (Sections \ref{sec:results:CNNB} and \ref{sec:results:singlesHPO}, panels A and B),
then we report the same model configurations but evaluated as ensembles (Sections \ref{sec:results:baseline-ensembles} and \ref{sec:results:singlesHPOensembles}, panels C and D).
In Section \ref{sec:results:ensemblesHPO} and panel E we report results of HPO scenarios that use ensembles within HPO. 
Lastly, Section \ref{sec:results:cheatingModel} describes a lower-bound, shown in panel F.

Post-hoc analysis in Section \ref{sec:posthoc:generalizability} describes how well validation set error correlates to test set error. And, finally, Section \ref{sec:posthoc:baseline} discusses the efficacy of HPO as compared to an expert.

\subsection{\CNNB{} Variance}
\label{sec:results:CNNB}
First, we obtain the distribution of error scores
of the baseline, \CNNB{},
by training 2000 times with random initializations. 
Scores are measured in RMSE, whose units match the prediction target: in this case, percent of dry matter (DM) content.
The distribution is shown in Figure \ref{fig:HPO_boxplot_groups} in panel A,
it has a mean of 0.843, 95\% CI [0.789, 0.896].
This means that if this model were deployed, we would expect the prediction error to be between 0.789 and 0.896 95\% of the time.
Mishra \& Passos \cite{Mishra2021mangoes} reported 0.79 for \CNNB{}.
One can reproduce a particular training run by seeding the random functions, but this only holds as long as nothing else changes---if you change the training data to include the test set (as you would in preparation for deployment), the resulting weights will again converge to a random setting.
%

\subsection{\SinglesHPO{} Evaluations Without Ensembling}
\label{sec:results:singlesHPO}

\begin{figure}[tb!]
    \centering
    \includegraphics[keepaspectratio,width=1\linewidth]{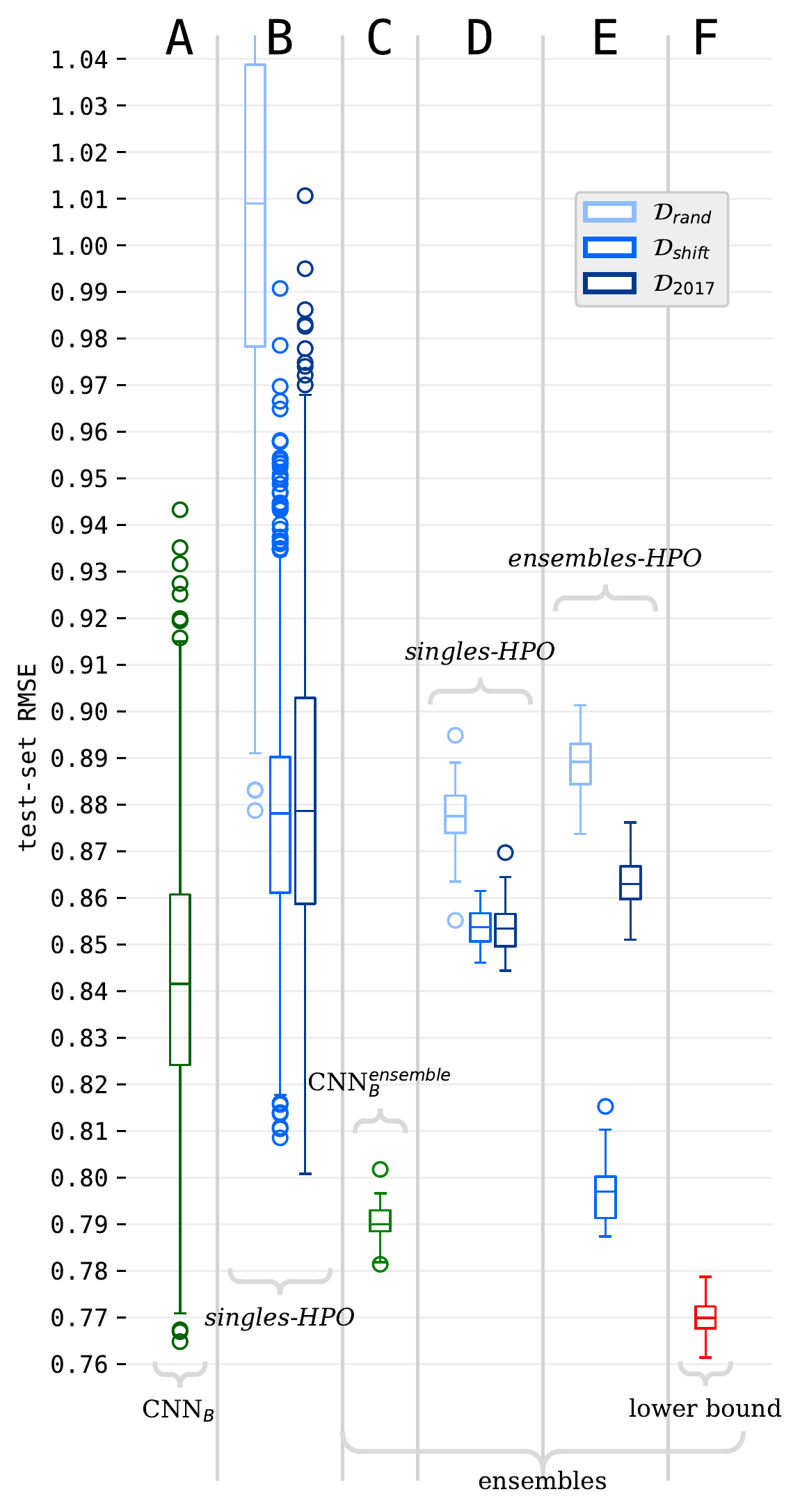}
    \caption{
        Boxplot where each box a distribution over RMSE scores.
        The results are divided into 6 panels (labelled A--F).
        A is explained in Section \ref{sec:results:CNNB},
        B in \ref{sec:results:singlesHPO},
        C in \ref{sec:results:baseline-ensembles},
        D in \ref{sec:results:singlesHPOensembles},
        E in \ref{sec:results:ensemblesHPO},
        and F in \ref{sec:results:cheatingModel}.
        The top and bottom of the boxes are the upper and lower quartiles of the distribution
        with a line at the median. 
        The whiskers extend from the box to show the range of the data and
        the points beyond the whiskers are considered outliers.
        %
    }
    \label{fig:HPO_boxplot_groups}
\end{figure}

Figure \ref{fig:HPO_boxplot_groups}, panel B, shows the results of the first set of hyperparameter optimization runs, \singlesHPO{}, 
with each partitioning of the dataset (\D{rand}, \D{shift}, and \D{2017}).
After obtaining the best neural network configuration from HPO (line \ref{line:eval-HPO} of Figure \ref{code}),
this neural network is 
trained on the original (\colorDrand{\D{rand}}) partitioning of the dataset (line \ref{line:eval-retrain}) 
and evaluated on the test set (line \ref{line:eval-score});
lines \ref{line:eval-retrain}-\ref{line:eval-score} are repeated 2000 times to obtain a distribution of RMSE scores.
The resulting distributions are visualized in the boxplot (Figure \ref{fig:HPO_boxplot_groups}, panel B).

HPO on \D{rand} was least successful (RMSE of 1.01 on average);
\D{shift} and \D{2017} fared better with a mean around 
0.88.
This supports the original hypothesis that HPO overfits to \D{rand}[val] 
and that forcing HPO to extrapolate---by validating on samples in \D{shift}[val] or \D{2017}[val]---results 
in a model that is better able to extrapolate to future samples.
\D{shift} and \D{2017} HPO scenarios performed similarly, except \D{2017} had higher variance as expected because it was trained on fewer examples.
%
%
Next, we look at using ensembles to reduce variance.

\subsection{\CNNB{} Evaluation With Ensembling}
\label{sec:results:baseline-ensembles}

\begin{figure*}[htb]
    \centering
    \includegraphics[keepaspectratio,width=\linewidth]{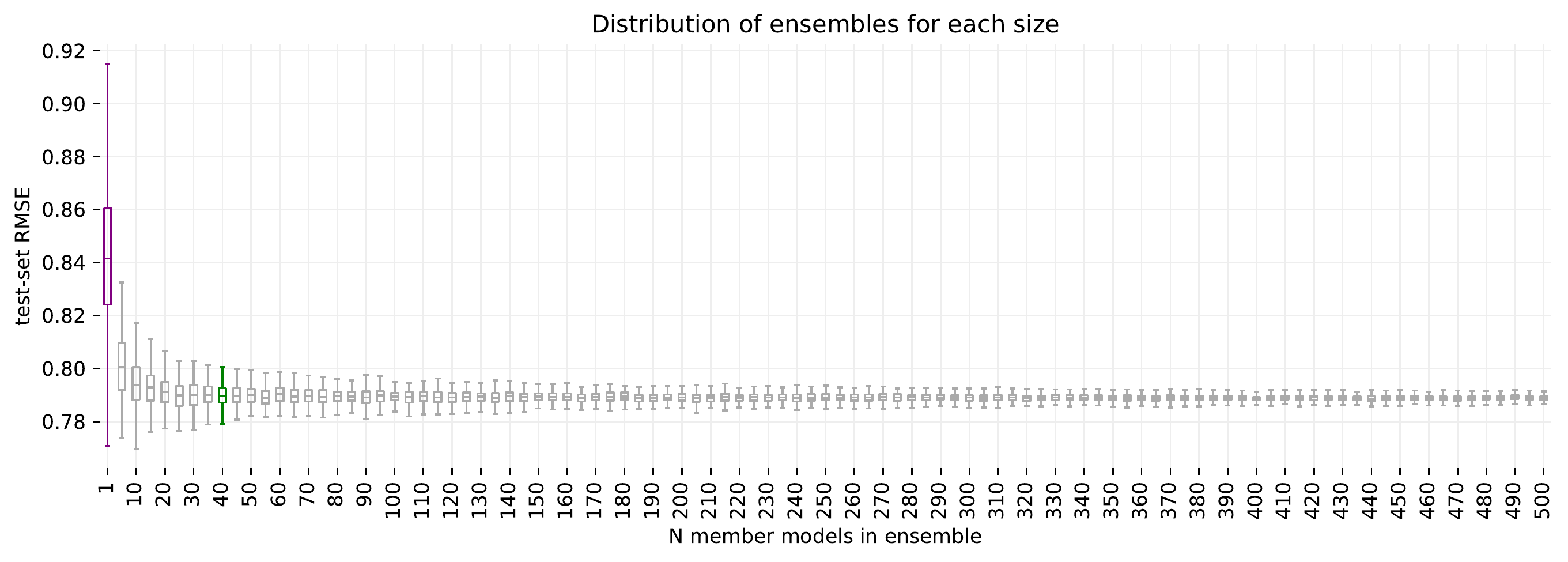}
    \caption{RMSE of test set using \CNNB{} model as a function of ensemble size. 
             Size $N=1$ (\textcolor{violet}{purple}) indicates the distribution of RMSE scores over 2000 single re-trainings (with random initialization) of the model.
             Each boxplot (where $N>1$) is a distribution over 200 simulated ensembles, where each ensemble is formed from a unique combination of trained models (from a pre-computed set of 2000).
             We use ensembles with 40 member models in our experiments (\textcolor{mpl_green}{green}).}
    \label{fig:ensemble_size}
\end{figure*}

Before doing ensemble experiments,
we need to decide how many models to include in the ensemble.
Since each member of the ensemble must be trained, more computation is required for each additional model,
though they are readily parallelizable if parallel computing resources are available.
The number of members in the ensemble can be chosen based on the trade-off between required accuracy and available computation.
Using the baseline, \CNNB{}, we compute the distribution of scores
for single models ($N=1$) up to ensembles with 500 member models ($N=500$),
shown in Figure \ref{fig:ensemble_size}.
The $N=1$ distribution is also the distribution for \CNNB{} reported in Figure \ref{fig:HPO_boxplot_groups}, panel A.

The distributions in Figure \ref{fig:ensemble_size} are computed as follows.
We first trained 2000 single models using the baseline configuration (\CNNB{}).
Then we simulated ensembles of every size, $N$, from the pre-computed set of 2000.
Each ensemble is a random, but unique, combination of $N$ models.
We generate 200 ensembles for each $N$ and plot the distribution of RMSE scores.
    
From Figure \ref{fig:ensemble_size} we observe a sharp drop in variance and test-set error for ensembles of size $N=5$.
Variance continues to decrease with greater $N$ but tapers off.
The median error drops quickly at first, but quickly converges.
We use 40 ensemble members in our experiments which is when 
the distributions of errors start to plateau while keeping computation reasonable based on our available compute resources.
An ensemble of size 40 improves the mean error to 0.790, 95\% CI [0.782, 0.799];
we refer to this model as \CNNBensemble{}.
We compare the results that follow to \CNNBensemble{},
whose distribution of errors is shown in Figure \ref{fig:HPO_boxplot_groups}, panel C.

\subsection{\SinglesHPO{} Evaluations With Ensembling}
\label{sec:results:singlesHPOensembles}

Since ensembling \CNNB{} worked well,
we also test the best configurations from each \singlesHPO{} scenario as ensemble models.
In the $\var{final\_evaluation}$ function (Figure \ref{code}),
the best hyperparameter configuration from HPO (line \ref{line:eval-HPO}) 
is re-trained 40 times (repeat line \ref{line:eval-retrain} 40 times).
Predictions from the 40 models are averaged to make the ensemble.
In line \ref{line:eval-score}, RMSE is computed using the ensemble's predictions on the test set (\D{}[test]).

Prediction error from ensembles also vary due to randomness, 
so we repeat the process to obtain and report the distribution of RMSE scores.
Using the 2000 models from Section \ref{sec:results:singlesHPO},
we group them into groups of 40 models
which makes for 50 ensembles (50 $\times$ 40 is 2000).
The resulting distributions are shown in Figure \ref{fig:HPO_boxplot_groups}, panel D.
The results show that the variance has tightened substantially
and the error on average has also dropped significantly.
However, \CNNBensemble{} remains better than all of these;
the next experiment is to see if ensembling \textit{within} hyperparameter optimization leads to an improvement.

\subsection{\EnsemblesHPO{} Evaluations}
\label{sec:results:ensemblesHPO}
The second set of hyperparameter experiments 
use ensembles for each configuration in HPO \textit{and} to evaluate the final configurations.
Results are shown in Figure \ref{fig:HPO_boxplot_groups}, panel E.
For each configuration in HPO (see $\var{HPO}$ function, line \ref{line:HPO-BO}),
neural network training is repeated 40 times (repeat line \ref{line:HPO-train-NN}).
Then, on line \ref{line:HPO-score}, the ensemble's RMSE is calculated.
The $\var{final\_evaluation}$ function proceeds with ensembling as well,
just as described in the previous section (Section \ref{sec:results:singlesHPOensembles}).
The resulting distributions of 50 ensemble models are visualized in the boxplot (Figure \ref{fig:HPO_boxplot_groups}, panel E).

The best \ensemblesHPO{} result was obtained using the \D{shift} partitioning and results in a mean RMSE of 0.796, 95\% CI [0.784, 0.809].
This distribution closely overlaps with the \CNNBensemble{} distribution.
This suggests that \CNNB{} is already very good and, indeed, the authors did a thorough investigation into finding a model that would generalize;
ensembling this model makes it even better.
The combination of \D{shift} with \ensemblesHPO{} achieved competitive performance to \CNNBensemble{}
and
this is achieved completely automatically without expert guidance.

When switching from \singlesHPO{} to \ensemblesHPO{}, a significant improvement is seen with \D{shift} whereas \D{2017} became worse.
The difference between \D{shift} results and \D{2017} results may be due to the difference in size of the calibration sets: 
\D{shift}[cal] is bigger than \D{2017}[cal] (sizes are listed in Table \ref{tbl:val-sets}).
While training networks during HPO, 
networks trained using \D{shift}[cal] see more data than when using \D{2017}[cal];
this may provide a better estimate of performance
which may lead HPO to a better final configuration.

As for the improvement from \singlesHPO{} to \ensemblesHPO{},
this may be explained by the reduced variance.
Since ensembles reduce variance, 
the ensemble members can individually exhibit larger variance than single models would allow.
That is, individual networks are given more leniency to overfit more or be less regularized because the ensemble reduces the added variance while keeping the error low.
HPO then picks a configuration whose \textit{ensemble} error is optimized, rather than the error of an individual model, 
leading to the best-possible ensemble.

\begin{figure*}[htbp]
    \centering
    \includegraphics[keepaspectratio,width=\linewidth]{{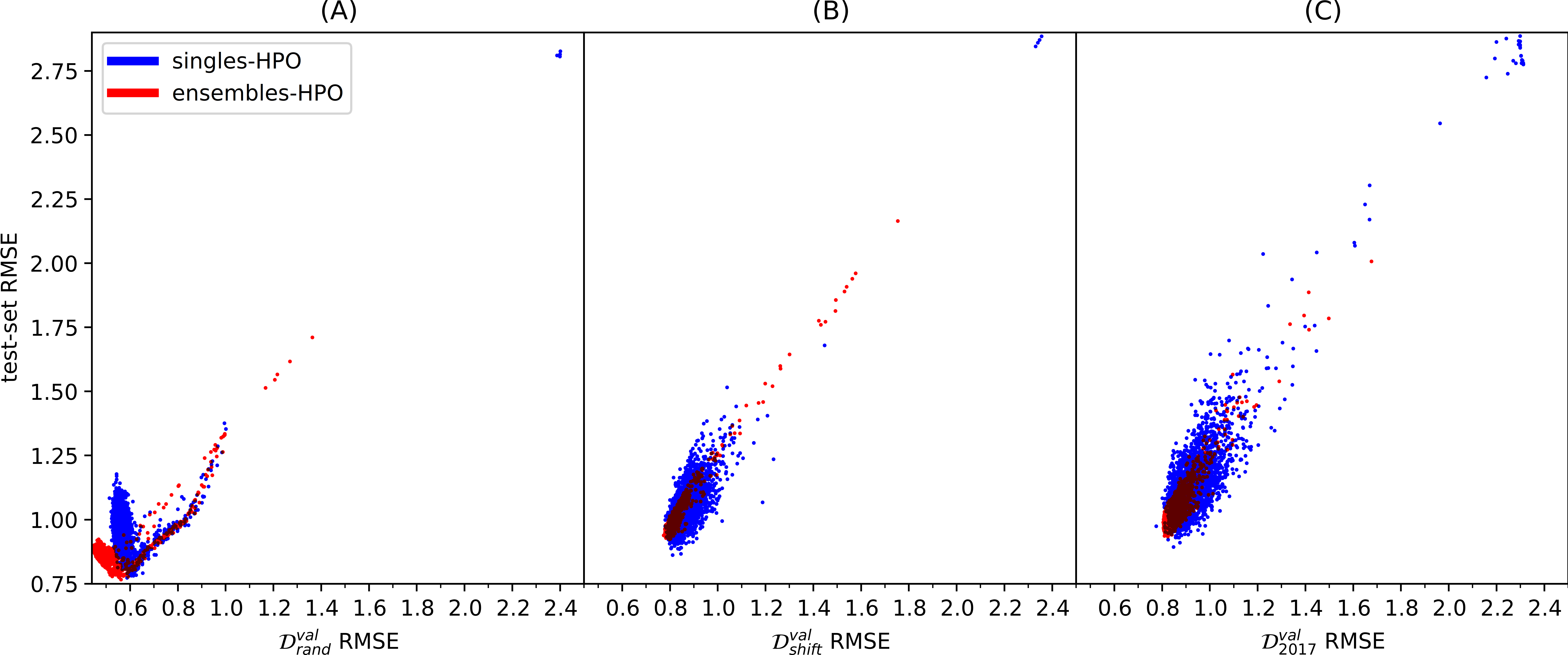}}
    \caption{Best viewed in color. Relationship between RMSE on the test set versus validation set for each HPO scenario.
             %
             \textcolor{mpl_blue}{Blue} points: HPO configurations trained once (\singlesHPO{} scenarios).
             \textcolor{mpl_red}{Red} points: HPO configurations trained as an ensemble of 40 models (\ensemblesHPO{} scenarios).
             Plot is cutoff at 2.5 on the x-axis and 2.9 on the y-axis for readability 
             (14 points from \D{2017} (\singlesHPO{}) are cutoff but are positioned where you'd expect by linearly extrapolating to $(3.34, 4.40)$).
             %
            }
    \label{fig:loss_cor}
\end{figure*}

\subsection{Generalizability of Validation Set}
\label{sec:posthoc:generalizability}

Hyperparameters are chosen to minimize validation error, with hopes that the hyperparameter configuration also minimizes test set error.
This assumes that validation set error is correlated to test set error \cite{cawley2010overfitting};
here we check whether this is achieved.

Figure \ref{fig:loss_cor} shows validation-set and test-set error for all configurations tried during HPO
and for all 6 HPO scenarios.
Each point is the result of training one hyperparameter configuration on the calibration set then evaluating it on validation and test sets.
There are very few points with high validation-set or test-set error
because HPO focuses its search on configurations with low validation-set error.
Validation-set error and test-set error are worse for the \D{shift} and \D{2017} HPO scenarios, as expected, 
simply because these neural networks don't get to train on any samples from harvest season 2017.
Networks trained using \D{rand} have the advantage of getting to see samples from harvest season 2017,
which are useful for making predictions on harvest season 2018 (the test set).
Three important observations can be made from this figure:

\begin{enumerate}
    \item 
        The most important observation is the ``V'' shape in Figure \ref{fig:loss_cor} (A) (\D{rand} with \singlesHPO{} and \ensemblesHPO{});
        as HPO progresses toward lower validation-set error the test-set error drops at first, but then increases again.
        This indicates overfitting.
        Similar to how neural network weights can overfit,
        overfitting in model selection is a significant problem \cite{cawley2010overfitting}.
        %
        %
        The two HPO scenarios in Figure \ref{fig:loss_cor} (A) 
        exhibit overfitting likely because the validation and calibration samples are randomly partitioned
        which means that for every sample in the validation set, a similar sample in the calibration set can be found.
        Fitting the calibration set well (or overfitting it) results in a good fit of the validation set too.
        %
        %
        However, since the test set is an entire separate season, it is not as similar to the calibration samples as the validation set is,
        suggesting that the models are overfitting to seasonal characteristics in the data.
        %
        %
        When partitioned across time, as in \D{shift} and \D{2017} shown in Figure \ref{fig:loss_cor} (B) and (C), the relationship between test and validation error is more linear and monotonic.
        In these, minimizing validation set error is a good proxy for minimizing test set error,
        leading to a model that doesn't overfit the past and can extrapolate better to the future.

    \item
        Secondly, the cloud of points is tighter for all \ensemblesHPO{} scenarios (red points in Figure \ref{fig:loss_cor}) than for \singlesHPO{}.
        This shows that ensembling scenarios have better correlation between validation and test error.
        Furthermore, comparing \D{shift} to \D{2017}, we see in Figure \ref{fig:loss_cor} (B) and (C) that
        the cloud of points is tighter for \D{shift},
        indicating that \D{shift} with \ensemblesHPO{} has better correlation.
    
    \item
        Lastly, the points in \D{shift} with \ensemblesHPO{} appears to be the most correlated, and least spread out, across all the scenarios 
        (red points in Figure \ref{fig:loss_cor} (B)),
        which may explain why this scenario performed the best in the final evaluation (see Figure \ref{fig:HPO_boxplot_groups}, panel E).
\end{enumerate}

\subsection{Baseline Compared to HPO Configurations}
\label{sec:posthoc:baseline}

In this section we consider how the baseline (\CNNB{}) configuration would have performed had it been a candidate configuration during HPO.
Specifically, we look at validation-set error of \CNNB{} compared to all the validation scores seen during two HPO scenarios: \D{rand} with \singlesHPO{} and \D{shift} with \ensemblesHPO{} (the worst and best scenarios).

\begin{itemize}
    \item 
    Firstly, for \CNNB{} trained using \D{rand}, 
        validation error is 0.609 on average. 
        Compared to \D{rand} with \singlesHPO{} (blue points in Figure \ref{fig:loss_cor} (A)), 
        this error is approximately at the bottom of the ``V'' shape.
        This is not the minimum validation-set error, so \CNNB{} would not have been chosen.
        It is, however, close to where test set error is minimized.
        Other configurations work better on the validation set.

    \item
    Secondly, for \CNNB{} trained using \D{shift} with ensembling (i.e. \CNNBensemble{}), 
        validation error is 0.792 on average.
        Compared to \D{shift} with \ensemblesHPO{} (red points in Figure \ref{fig:loss_cor} (B)),
        this validation-set error is close to the best HPO configuration which has a validation error of 0.774. 
        Since HPO picked a configuration whose validation set error is essentially on par with \CNNBensemble{},
        we can say that this HPO objective is successfully guiding the search towards configurations on par with the expertly-chosen one.
\end{itemize}

\subsection{Lower Bound}
\label{sec:results:cheatingModel}
From Figure \ref{fig:loss_cor},
the model configuration with the best test-set loss may give an idea of what the lower bound is
for our hyperparameter search space
and how far away our models are from it.
The hyperparameter configuration that minimizes test-set error is from
\D{rand} with \ensemblesHPO{} (Figure \ref{fig:loss_cor} (A)) with a test-set error of 0.766.
This configuration was not chosen by HPO, however, because its validation error was not the minimum.
We can't deploy this model because it likely overfits the test set.
Re-trained 50 times, it achieves a mean of 0.770 (for reference, \CNNBensemble{} has mean of 0.790)
and its error distribution is shown in Figure \ref{fig:HPO_boxplot_groups}, panel F.
This gives a lower bound on what we could possibly expect to achieve with further improvements.

\section{Conclusion}

The goal of this study was to elucidate hyperparameter tuning methods for building neural networks that extrapolate to future samples.
Experiments showed that HPO overfits the validation set when the validation set is randomly selected.
We found that HPO worked better when trial 
configurations were validated on the latest samples (either the latest 1/3 or a whole harvest season) 
rather than random samples.
For these validation sets, validation-set error better correlates to test-set error,
thus guiding hyperparameter optimization to a model that extrapolates well.

Upon re-initializing and re-training, neural networks produce varying prediction errors.
Ensembles of randomly-initialized neural networks reduced the variance and average error substantially for all HPO models and the previous state-of-the-art.
Lowering the variance of the neural networks during HPO---combined with the new validation set---%
led to considerable improvement for hyperparameter optimization
which found a hyperparameter configuration on par with \CNNBensemble{}.

For practitioners building neural networks for spectroscopy, we make three recommendations as a result of this study:
(1) Optimize hyperparameters by validating on 
samples that correspond to how the model will ultimately be used.
If the model should be expected to extrapolate to next year,
then samples in the HPO validation set should also be a year separate from data used to train the neural network.
Otherwise, as we found, hyperparameters overfit in a way similar to traditional model overfitting.
(2) Reduce neural network variance and improve accuracy using ensembling, such as ensembles obtained by re-initializing the neural network many times, re-training it, and averaging all the models' predictions.
(3) Consider using ensembles both in deployment and in validating HPO configurations.




\bibliography{references}

\end{document}